\documentclass[12pt]{article}
\usepackage{graphicx}
\usepackage{amsfonts}
\usepackage{amscd}
\usepackage{latexsym}
\usepackage{epsf}

\newcommand{\be}{\begin{equation}}
\newcommand{\ee}{\end{equation}}
\newcommand{\bea}{\begin{eqnarray}}
\newcommand{\eea}{\end{eqnarray}}

\newcommand{\nn}{\nonumber}
\newcommand{\fr}{\frac}

\newcommand{\al}{ \alpha }

\newcommand{\ep}{ \epsilon }

\newcommand{\ra}{\rightarrow}

\usepackage[latin9]{inputenc}
\usepackage{textcomp}
\usepackage{amsmath}
\usepackage{multirow}
\usepackage{color}
 
\newcommand {\si} {\sigma} 
\newcommand {\vep} {\varepsilon} 

\newcommand {\pb} {\bar{p}} 
\newcommand {\Pb} {\bar{P}} 
\newcommand {\Ybar} {\bar{Y}}

\begin{document}
\title{\bf Influence of Specific Surface Area of  Powder on Hydrogen Desorption Kinetics for  Metal Hydrides}
\author{
{\sc I.~V.~Drozdov}\thanks{Corresponding author, e-mail: drosdow@uni-koblenz.de},
 }
\maketitle

\begin{abstract}

 The observable results for desorption kinetics by powder of metal hydride 
on the example of mangesium hydride are reproduced with the model
formulated in terms of specific surface of powder.
A volumetric measurement of hydrogen desorption process is evaluated on an example of wet ball milled magnesium hydride, and can be applied generally for any metal hydride. 
 The exact solution of the model reproduces the shape of experimental curves for desorption process providing a satisfying agreement with experimental data.
\end{abstract}

\small
\section{Introduction}

 Metal hydrides are considered as a potential hydrogen storage material because they have 
 a high storage capacity by weight. Especially the magnesium hydride reaches the theoretical maximal value of 7.66 wt\%, but its main drawback are the high sorption temperature (573-673K) and the sluggish sorption kinetics, commonly typical for metal hydrides (of light metals in the first instance). We restrict the following discussion on this material only, retaining that the modelling outlined below is nonspecific and can be applicable for any metal hydride.

In recent years, significant process has been made using nanocrystalline Mg hydride
produced by high energy milling and adding suitable catalysts in order to improve the
sorption kinetics. Without catalysts the desorption temperature of high energy milled
Mg hydride is still higher than 573K. 

As it was reported from the recent research (\cite{meng}), wet ball milling method was used to produce nanocrystalline Mg hydride which is different from the conventional high energy ball milling. As we know, during high energy milling, particle size is decreased
significantly, that influence the sorption behavior.

 Additionally, it has been found, that 
 the wet ball milled powder demonstrates a better desorption kinetics as the same dry ball milled powder with the same particle size. The reason could be probably the higher  specific surface area of the former one. Hence a model operating with the specific surface area instead of particle size would be more suitable for evaluation of sorption capability of powders.

 \section{Model reformulation in terms of {morphofactor} and specific surface area}
\subsection{ The morphological factor of particles }

 A transport of hydrogen in-/outwards of a metal particle with the characteristic size $d$
 occurs through its surface. The rate of the total hydrogen permeation is therefore proportional 
to the surface area $S$, i.e. $\sim d^2$, whereat the hydrogen content of the
 particle is proportional to its volume $V$, $\sim d^3$. From these suggestions it is to expect the characteristic
 sorption time $\tau$ (especially for short time distances) to be dependent linearly on the characteristic particle
 size $d$. The dimensionless proportionality factor $1/\xi:=\sqrt[3]{V}/\sqrt{S}$ is dependent on geometry/morphology of the particle.

 For strongly convex particles it lies typically between 2.684 (tetrahedron) and 2.199 (ball). For
 concave poly-particles with cavities (e.g. wet ball milled) or plain cakes (like original $MgH_2$ particles),
 this factor is to expect overcoming a double value.  
 It means that at the same particle size, the specific surface of the compound can differ up to several times, provided
 by the surface morphology, that can influence the characteristic sorption time.

  The dependence outlined above should hold in the leading order if we compare powders with different particle sizes. 
 The characteristic sorption times of particles 1 and 2 are expected to relate to each other like their 
 characteristic sizes,
 \be \tau_1/\tau_2 \sim d_1/d_2. \label{dimension}\ee
 It should be the main effect providing an advantage
 of wet ball milled compound over the dry ball milled and other conventional ones. 
 This hypothesis proposed in our recent publication \cite{1stpaper}, devoted to the simple sorption modelling, 
would be subjected an experimental validation if we have samples consisting of equal sized particles with the same
morphology, e.g. spherical. In fact an attempt to verify the relation (\ref{dimension}) for different kinds of materials,
e.g. as-received and dry-ball milled, stumbles on the deviation up to several times. As possible reasons the next obvious
reasons can be considered:

1. all particles in a sample are of different sizes, described by a certain size distribution; 

2. samples of different materials consist of particles with different morphology, the ratio of morphofactors $\xi_1/\xi_2$ can reach several times, as noted above.
  
  The exact measurement of size distribution is extensive and the evaluation of the results is not unique, while the measurement of specific surface via BET is successfully available and is exact enough. Instead of the size distribution together with the morphofactor $\xi$ we can take this specific surface area for a governing parameter.
  
The idea of the improvement of the model proposed below is
 to take the specific surface into account instead of the characteristic particle size as it was considered before \cite{1stpaper}. 
Additionally, the shrinkage of the outer surface caused by the volume shrinkage is considered, and the influence of this effect
 is accounted and estimated in the model. The analysis of results is simplified thereby, that the model still allows an analytical approach.

\subsection{ Improvement and generalization of the linear model }

\begin{minipage}[b]{8.5cm}
 The spherical symmetry of particles as considered in the previous formulation \cite{1stpaper} of the model,
 was assumed there following a number of similar models \cite{castro, gabis}
 only for the sake of transparency. 
 
In fact, the confluent model \cite{gabis} under consideration does not demand any symmetry: the desorption
 rate depends only on the total surface of the particle and 'does not see' the surface of the inner $\beta$-core, 
since the $\al$-concentration between these two surfaces remains constant homogeneous due 
 to fast $\al$-diffusion as assumed. 
 So, the results can be performed  for a particle of an arbitrary form, without further requirements.
 \end{minipage}
\vspace*{1cm} \includegraphics[scale=0.22]{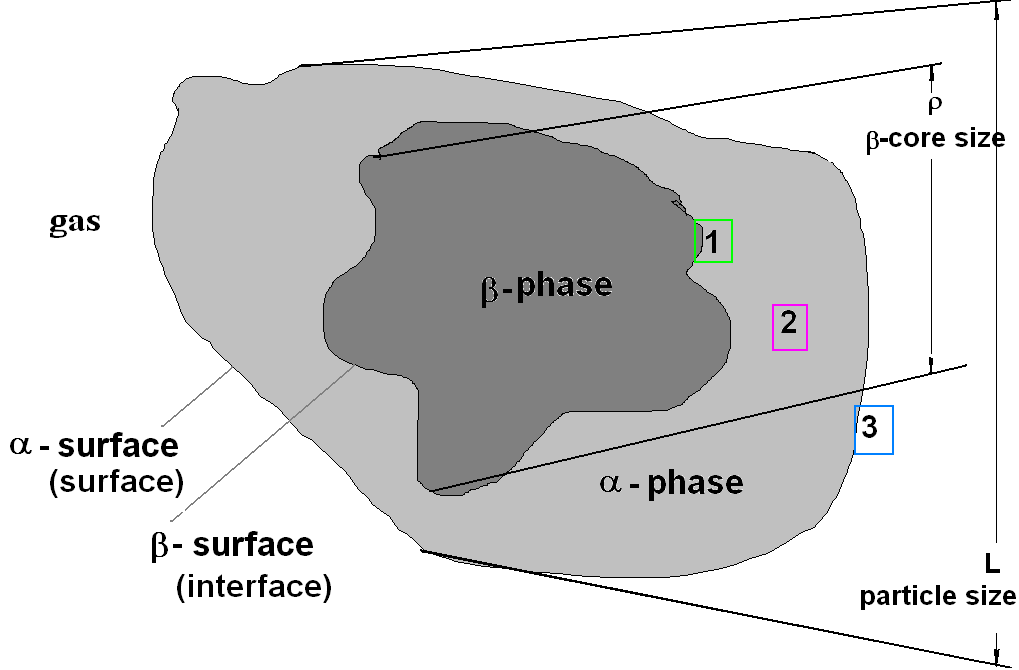}
 We remind here briefly on the formulation of the model \cite{1stpaper}:
The inner core consists of the stoichiometric $MgH_2$ in the $\beta$-phase ($\beta$-core), having 
a total volume $v_{\tiny\beta}$ shrinking during the desorption. 
In fact, the geometry and deformation of the $\beta$-
core during the desorption as well as the number of such $\beta$-cores in a single particle is unimportant 
for results of the model. The remaining space of the particle is the $\al$-phase consistent of dissolved $H^+$ ions in 
the metallic magnesium with the constant (for the given temperature) molar concentration $X, \ [mol/m^3]$.
 The molar concentration $Y$ of hydrogen atoms in the $\beta$-phase is always constant, $Y=110119 \ [mol/m^3]$  

Then we have, for the balance of desorbed hydrogen atoms $\dot{\nu},\ [mol/s]$ 
\be
\dot{\nu}= -(Y-\eta X) \fr{d v_\beta}{d t},
\label{nu_rate}
\ee
with initial and final conditions: 
\be v_\beta (t=0)=v_0;\ \ \ v_\beta (t=\tau)=\eta v_0, \ee
$v_0$- initial volume of the single particle, $\tau$- the life time of complete decay of the $\beta$-phase.
 The volume shrinkage coefficient $\eta$ is taken into account, because of different densities of magnesium
 hydride and metallic magnesium \cite{1stpaper}.
Then, for the current volume of the particle we have 
\be
v(t)= (1-\eta)v_\beta(t)+\eta v_0
\ee

 The surface $s$ of the particle is in any time $t$ related to the volume by 
\be
s(t)=\fr{1}{\xi^2} v(t)^{2/3}= \fr{1}{\xi^2} \left[  (1-\eta) v_\beta(t)+\eta v_0  \right]^{2/3}.
\ee
 Finally, the desorption kinetics is controlled by two surface parameters of the reaction $2 H \rightleftharpoons H_2$,
the desorption constant $b$ and the re-adsorption $k$, whereat the desorption rate is proportional to the 
particle surface $s$:   
\be
\dot{\nu}_i=s(b X^2- k p), 
\ee
$p$-total outer pressure of molecular hydrogen.

With $p_i$ -the partial pressure produced by desorption from i-th particle in the volume $V$ we have
\be
\nu_i(t)=(v_0-v_\beta(t))[ Y-\eta X ]=\fr{2}{R} \{ V/T \} p_i(t)
\ee
and a differentiation of this relation combined with \ref{nu_rate} provides
\be
\fr{2}{R}  \{ V/T \}\dot{p}_i=-\dot{v_i}_\beta [ Y-\eta X ] = \fr{b X ^2 -k p}{\xi^2} 
\left[  (1-\eta) v_\beta +\eta v_0 \right]^{2/3}, 
\ee
here 
\be
p=\sum\limits_i p_i
\ee

Now, introducing the notations 
\be A(p)= \fr{R}{\xi^2}\cdot \fr{b X^2-k p}{ 2\{ V/T \}};\ \ \ B=\fr{2}{R} \fr{(\eta-1)\{ V/T \} }{Y-\eta X}\ee
we express the evolution of pressure  $p$ as measured: 
\be
\dot{p}_i= A [B p_i + v_{0i} ]^{2/3}
\ee
 
The equation cannot be simply summarized over $p_i$ to obtain the total pressure
$p$, because of the power 2/3. 

 We can verify this formula first for the special case of equal particles. To this end we assume the powder sample
 to contain $N$ particles of equal size and equal form (morphology). It means then    
\be p=N p_i,\ \ \ \bar{s}=N s_i,\ \ \ \bar{v}=N v_{0 i}\ee
for the total pressure, total surface of desorption and the total initial volume of powder in the sample,
whereat $\bar{v}$ and $\bar{s}$ are related by 
\be
(\bar{v}/N)^{1/3} = \xi ( \bar{s}/N)^{1/2}.
\ee
  The factor $\xi$ can be generally established using indirect measurements. 
 In this special case we prefer instead of $\xi$ other parameters measured directly to perform the calculation to
 compare with experimental results.

 The knowledge of the sample mass $m$, mass density $\varrho$ and the specific surface $\si$ (BET) allows the
 elimination of the morphological factor $\xi$ by
\be 
\fr{m}{\varrho N} = \xi^3 \left(  \fr{\si m}{N} \right)^{3/2}
\ee

The resulting evolution of the total pressure is described by

\be
\dot{p}=\fr{\si m R}{ 2\{V/T\} } (b X^2- k p)
\left[  \fr{2\varrho }{m R} \cdot \fr{ (\eta-1)\{ V/T\} }{Y-\eta X} p
 + 1 \right]^{2/3}
\label{kinetic}
\ee
where two terms in brackets are of magnitude comparable to each other.
 The desorption measurements have been carried out for several samples with the order of magnitudes:\\
 $1-\eta=0.23, \ \ \varrho=1450\ kg/m^3,\ \  \{V/T\}\sim 10^{-7} m^3/K, \ \ m\sim 100 mg,\ \
 \ Y-\eta X\sim 10^5\ mol\ H/m^3$ 
  
 It provides the first term in brackets about 0.08 compared to unity. Therefore, at the beginning of desorption 
(for small $p$), the desorption kinetics is quite well described by the simplified equation:
\be
\dot{p}=\fr{\si m R}{2 \{ V/T \}} (b X^2- k p)\equiv \vep_{\si,m} (b X^2 - k p)
\label{kinetic_simpl}
\ee 
Especially for the case that all particles are initially of the spherical form, $ \si =3/(L\varrho) $
and we obtain the kinetic formula of \cite{1stpaper}.
 However, if the pressure increases e.g. doubled (trebled), the first term in brackets (\ref{kinetic}) 
becomes 0.16 (0.24) respectively, and in principle may not be neglected anymore.  
 
 Finally, the description in terms of specific surface can be subjected to verification, under 
assumption, the kinetic desorption and re-adsorption constants $k$ and $p$ as well as the critical
$\al$- concentration $X$ are inherent properties of the material, independent of geometry/morphology
 and remain therefore the same for all kinds of compound (original, dry- and wet-ball-milled).      

We take the solution of (\ref{kinetic_simpl}) in a linear approximation of kind
\be p(t)=\fr{b X^2}{k} \left( 1- e^{-\vep_{\si,m}k t } \right)\sim  {b X^2} \vep_{\si,m} t \ee
 The proportionality expected to hold for two different kinds 1 and 2 of compound will be :
\be
\fr{p_1}{p_2} = \fr{ m_1 {\cal P}_1 t_1   }{ m_2 {\cal P}_2 t_2 }\cdot \fr{\si_1}{\si_2}
\ee 
\begin{minipage}[b]{5cm}
As an example, two desorption curves for desorption of as-received $MgH_2$ (m=140\ mg , purity ${\cal P}=0.85$)
and dry ball milled (m=149\ mg, ${\cal P}=0.91)$. The approximatively linear increase of pressure for both samples 
between 50 and 200\ kPa ($p_1=p_2$) lasts 35 and 73 sec respectively.
\end{minipage}
\vspace*{0.5cm} \includegraphics[scale=0.28]{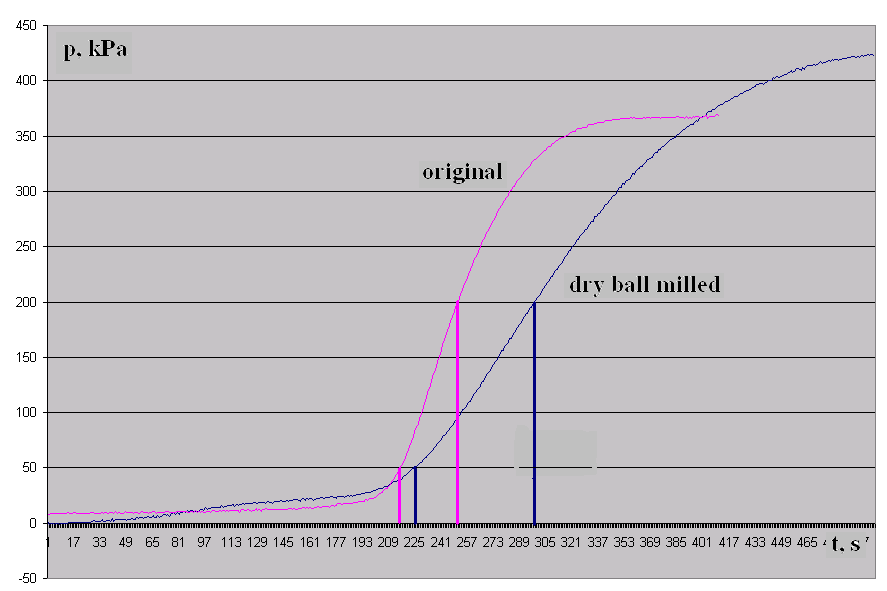}
It provides:
\be
\fr{ m_2 {\cal P}_2 t_2 }{ m_1 {\cal P}_1 t_1  }=2.376
\ee

 \subsection{ Analytical Solution }

 Introducing the notations: 
\bea 
\pb &=& \fr{b X^2}{k}\ \ \ \mbox{ for the 'threshold' pressure }\nn\\
\Ybar &=& \fr{Y-\eta X}{1-\eta}\ \ \ \mbox{ for the 'effective desorbable' molar concentration of hydrogen in compound }\nn\\
\ep &=& \fr{m R}{2 \{ V/T \}}\ \ \ \mbox{ for the experimental equipment factor, like $\vep$ \cite{1stpaper} } 
\eea 

in the (\ref{kinetic}), we rewrite it in the form
\be 
\fr{d p}{(\pb-p ) \left[ 1 - \fr{\varrho}{\ep\Ybar}p \right]^{2/3}  } =\si \ep k\ d t
\ee  

with the further notations 
\be
\Pi:=\Ybar \fr{\ep}{\varrho};\ \
\Pb:=(\Pi-\pb)^{1/3};\ \ \ P_0:=(\Pi-p_0)^{1/3},
\ee
the solution obeying the initial condition $p(t=0)=p_0$ reads \cite{gradshteyn}
\bea
-\si \ep k \fr{\Pb^2}{\Pi^{2/3}} t &=& 
\fr{3}{2} \ln \fr{(\Pi-p)^{1/3}-\Pb  }{P_0-\Pb} \left[\fr{\pb-p_0}{\pb-p}\right]^{1/3}-\\
&-& \sqrt{3} \arctan \sqrt{3} \fr{ (\Pi-p)^{1/3} -P_0   }{2\left( \fr{(\Pi-p)^{1/3} P_0}{\Pb} +\Pb 
 \right)+(\Pi-p)^{1/3}+P_0 }\nn
\label{solution}
\eea
that is now suitable for graphical evaluations.

 In the Fig.5 shown below, 
 the desorption kinetics, described by the present pressure-time law (\ref{solution}) outlined above,
depicted by the red line, is compared with the simplified law 
\be
t(p)=\fr{1}{\kappa}\ln\fr{\pb-p_0}{\pb-p},
\label{simple}
\ee   
obtained in \cite{1stpaper}, where the effective shrinking of the specific surface due to desorption is not
taken into account (green line). As expected, this feature leads to the slowdown of desorption at higher
 pressures. This effect is the appreciable, the less is the sample mass in the volumetric setup. \\ 

\vspace*{0.5cm}\hspace*{-1cm} \includegraphics[scale=0.46]{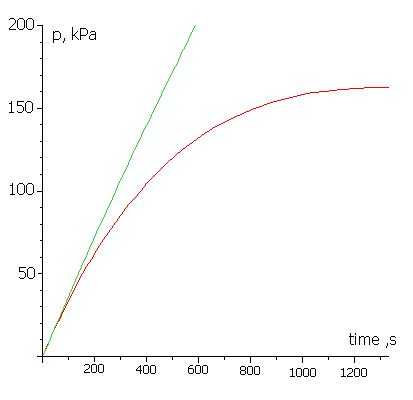}
\vspace*{0.5cm} \includegraphics[scale=0.45]{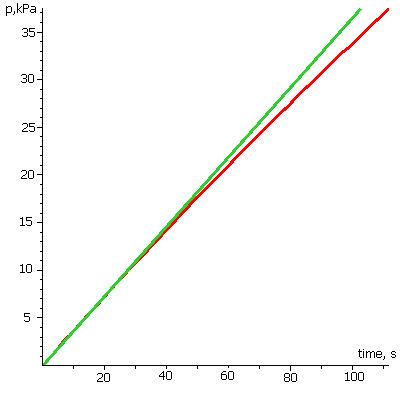}\\
{\bf \small
Fig.5 the deviation of kinetics (\ref{solution}) and (\ref{simple}) from each other for
the end of desorption(left); in fact these processes are only valid up to the pressure $p_{\al\ra\beta}$,
for 15 mg $MgH_2$ of 0.8 purity (case II of \cite{1stpaper}-complete desorption).  
}\\
\vspace*{0.1cm} \includegraphics[scale=0.58]{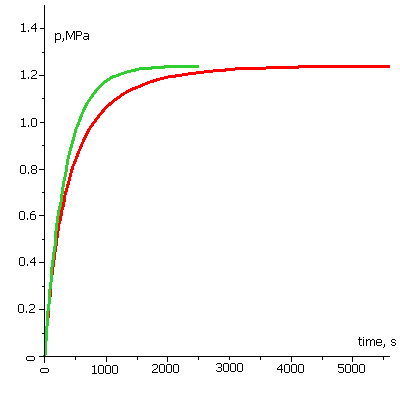} \\
{\bf \small Fig.6 Also, for the case I - reaching the threshold pressure $\pb$, (200 mg)are two kinetic laws different
}


\begin{thebibliography}{10}
\bibitem{meng}
Li~Meng,  PhD Thesis, Forschungszentrum Juelich (2010) 

\bibitem{1stpaper}
I.~Drozdov, Li~Meng, V.~Kochubey, R.~Va\ss en, G.~Mauer and D.~St\"over, submitted to Int.J.Hydr.Energy  

\bibitem{castro}
F.~J.~Castro, G.~Meyer,  J. Alloys. Compd. {\bf 330-332} (2002) 59-63

\bibitem{gabis}
I.~E.~Gabis, A.~P.~Voit, E.~A.~Evard, Yu.~V.~Zaika, I.~A.~Chernov, V.~A.~Yartys, J. Alloys. Compd.
 {\bf 404-406} (2005) 312-316

\bibitem{shapovalov}
V.~I.~Shapovalov, N.~P.~Serdyuk, O.~P.~Semik, Reprts. Acad. Sc. UkrSSR(ukr.) (1981) 99-101

\bibitem{popovic}
Z.~D.~Popovic, G.~R.~Piercy  Metall.Trans.A {\bf 6} (1975)1915-17

\bibitem{gradshteyn}
I.~S.~Gradshteyn, I.~M.~Ryzhik  {\it Tables of Integrals, Sums, Series \& Products } Moscow, (1963)     
\end{thebibliography}
\end{document}